\documentclass[12pt]{article}

\usepackage{graphicx}
\usepackage{amssymb}
\usepackage{amsmath}
\usepackage[toc,page]{appendix}

\usepackage[abbr]{jasa_harvard}    
\usepackage{JASA_manu}

\newtheorem{lemme}{Lemma}

\newtheorem{theorem}{Theorem}

\def\mbf#1{\mathchoice{\hbox{\boldmath $\displaystyle #1$}}
        {\hbox{\boldmath $\textstyle #1$}}
        {\hbox{\boldmath $\scriptstyle #1$}}
        {\hbox{\boldmath $\scriptscriptstyle #1$}}}

\begin{document}
\def\eop{\hfill\fbox{$\,$}}
\def\proof{\noindent{\bf Proof. \quad}}

\title{Approximate Bayesian Computation: a Nonparametric Perspective}

 \author{Michael G.B. Blum  \\
 CNRS, Laboratoire TIMC-IMAG, Facult\'e de m\'edecine, 38706 La Tronche \\
Universit\'e Joseph Fourier, Grenoble, France \\
 Phone: +33 (0)4 56 52 00 65\\
 email: \texttt{michael.blum@imag.fr}}

\date{}
\maketitle

\newpage
\begin{center}
\textbf{Abstract}
\end{center}
Approximate Bayesian Computation is a family of likelihood-free inference techniques that are well-suited to models defined in terms of a stochastic generating mechanism. 
In a nutshell, Approximate Bayesian Computation proceeds by computing summary statistics ${\bf s}_{obs}$ from the data and simulating summary statistics for different values of the parameter $\Theta$. The posterior distribution is then approximated by an estimator of the conditional density $g(\Theta|{\bf s}_{obs})$. In this paper, we derive the asymptotic bias and variance of the standard estimators of the posterior distribution which are based on rejection sampling and linear adjustment. Additionally, we introduce an original estimator of the posterior distribution based on quadratic adjustment and we show that its bias contains a fewer number of terms than the estimator with linear adjustment. Although we find that the estimators with adjustment are not universally superior to the estimator based on rejection sampling, we find that they can achieve better performance when there is a nearly homoscedastic relationship between the summary statistics and the parameter of interest. To make this relationship as homoscedastic as possible, we propose to use transformations of the summary statistics. In different examples borrowed from the population genetics and epidemiological literature, we show the potential of the methods with adjustment and of the transformations of the summary statistics. Supplemental materials containing the details of the proofs are available online.
\vspace*{.3in}


\newpage

\section{Introduction}

Inference in Bayesian statistics relies on the {\it full posterior distribution} defined as
\begin{equation}
\label{eq:bayes}
g(\Theta|D) = \frac{p(D|\Theta) \pi(\Theta)}{p(D)}
\end{equation}
where $\Theta \in \mathbb{R}^p$ denotes the vector of parameters and $D$ denotes the observed data. The expression given in (\ref{eq:bayes}) depends on the {\it prior distribution} $\pi(\Theta)$, the {\it likelihood function} $p(D|\Theta)$ and the marginal probability of the data $p(D)=\int_\Theta \! p(D|\Theta) \pi(\Theta) \, d\Theta$. 
However, when the statistical model is defined in term of a stochastic generating mechanism, the likelihood can be computationally intractable. Such difficulties typically arise when the generating mechanism involves a high-dimensional variable which is not observed. The likelihood is accordingly expressed as a high-dimensional integral over this missing variable and can be computationally intractable. Methods of inference in the context of these so-called {\it implicit statistical models} have been proposed by \citeasnoun{digglegratton} in a frequentist setting. Implicit statistical models can be thought of as a computer generating mechanism that mimics data generation. In the past ten years, interests in implicit statistical models have reappeared in population genetics where \citeasnoun{beaumontzhangbalding} gave the name of Approximate Bayesian Computation (ABC) to a family of likelihood-free inference methods. 

Since its original developments in population genetics \cite{fuli,tavarebaldinggriffithsdonnelly,pritchardetal,beaumontzhangbalding}, ABC has successfully been applied in a large range of scientific fields such as archaeological science \cite{wilkinstav}, ecology \cite{francoisetal,Jabot09}, epidemiology \cite{tanakafrancislucianisisson,blumtran}, stereology \cite{bortotetal} or in the context of protein networks \cite{ratmannetal07}. Despite the increasing number of ABC applications, theoretical results concerning its properties are still lacking and the present paper contributes to filling this gap.


In ABC, inference is no more based on the full posterior distribution $g(\Theta|D)$ but on the {\it partial} posterior distribution $g(\Theta|{\bf s}_{obs})$ where ${\bf s}_{obs}$ denotes a vector of $d$-dimensional summary statistics computed from the data $D$. The partial posterior distribution is defined as \cite{doksumlo}
\begin{equation}
\label{eq:partialpost}
g(\Theta|{\bf s}_{obs})=\frac{p({\bf s}_{obs}|\Theta) \pi(\Theta)}{p({\bf s}_{obs})}.
\end{equation}
Of course, the partial and the full posterior distributions are the same if the summary statistics are sufficient with respect to the parameter $\Theta$.%

To generate a sample from the partial posterior distribution $g(\Theta|{\bf s}_{obs})$, ABC proceeds by simulating $n$ values $\Theta_i$, $i=1\dots n$ from the prior distribution $\pi$, and then simulating summary statistics ${\bf s}_i$ according to $p({\bf s}|\Theta_i)$. Once the couples $(\Theta_i,{\bf s}_i)$, $i=1\dots n$, have been obtained, the estimation of the partial posterior distribution is a problem of conditional density estimation. Here we will derive the asymptotic bias and variance of a Nadaraya-Watson type estimator \cite{nadaraya64,Watson64}, of an estimator with linear adjustment proposed by \citeasnoun{beaumontzhangbalding}, and of an original estimator with quadratic adjustment that we propose. 

Although replacing the full posterior by the partial one is an approximation crucial in ABC, we will not investigate its consequences here. The reader is referred to \citeasnoun{lecam64} and \citeasnoun{abril94} for theoretical works on the concept of approximate sufficiency; and to \citeasnoun{joycemarjoram} for a practical method that selects informative summary statistics in ABC. Here, we concentrate on the second type of approximation arising from the discrepancy between the estimated partial posterior distribution and the true partial posterior distribution.

In this paper, we investigate the asymptotic bias and variance of the estimators of the posterior distribution $g(\theta|{\bf s}_{obs})$ ($\theta \in \mathbb{R}$) of a one-dimensional coordinate of $\Theta$. Section 2 introduces parameter inference in ABC. Section 3 presents the main theorem concerning the asymptotic bias and variance of the estimators of the partial posterior. To decrease the bias of the different estimators, we propose, in Section 4, to use transformations of the summary statistics.  In Section 5, we show some applications of ABC in population genetics and epidemiology .


\section{Parameter inference in ABC}

\subsection{Smooth rejection}

In the context of ABC, the Nadaraya-Watson estimator of the partial posterior mean $E[\Theta|{\bf s}_{obs}]$ can be written as

\begin{equation}
\label{eq:nw}
m_0=\frac{\sum_{i=1}^n \Theta_i K_B({\bf s}_i -{\bf s}_{obs}) } {\sum_{i=1}^n K_B( {\bf s}_i -{\bf s}_{obs} )}
\end{equation}
where $K_B({\bf u})=|B|^{-1}K(B^{-1}{\bf u})$, $B$ is the {\it bandwidth matrix} that is assumed to be non-singular, $K$ is a d-variate kernel such that $\int K({\bf u}) \, d{\bf u} =1$, and $|B|$ denotes the determinant of $B$. Typical choices of kernel encompass spherically symmetric kernels $K({\bf u})=K_1(\| {\bf u} \|)$, in which $\| {\bf u} \|$ denotes the Euclidean norm of ${\bf u}$ and $K_1$ denotes a one-dimensional kernel. To estimate the partial posterior distribution $g(\theta|{\bf s}_{obs})$ of a one-dimensional coordinate of $\Theta$, we introduce a kernel $\tilde{K}$ that is a symmetric density function on $\mathbb{R}$. Here we will restrict our analysis to univariate density estimation but multivariate density estimation can also be implemented in the same vein. The bandwidth corresponding to $\tilde{K}$ is denoted $b'$ ($b'>0$) and we use the notation $\tilde{K}_{b'}(\cdot)=\tilde{K}(\cdot/b')/b'$. As the bandwidth $b'$ goes to $0$, a simple Taylor expansion shows that
$$
E_{\theta'}[\tilde{K}_{b'}(\theta'-\theta)|{\bf s}_{obs}] \approx g(\theta|{\bf s}_{obs}).
$$
The estimation of the partial posterior distribution $g(\theta|{\bf s}_{obs})$ can thus be viewed as a problem of nonparametric regression. After substituting $\Theta_i$ by $\tilde{K}_{b'}(\theta_i-\theta)$ in equation ($\ref{eq:nw}$), we obtain the following estimator of $g(\theta|{\bf s}_{obs})$ \cite{rosen69}
\begin{equation}
\label{eq:est1}
\hat{g}_0(\theta|{\bf s}_{obs})=\frac{\sum_{i=1}^n \tilde{K}_{b'}(\theta_i-\theta) K_B({\bf s}_i -{\bf s}_{obs})} {\sum_{i=1}^n K_B({\bf s}_i -{\bf s}_{obs})}.
\end{equation}
The initial rejection-based ABC estimator consisted of using a kernel $K$ that took 0 or 1 values \cite{pritchardetal}. This  method consisted simply of rejecting the parameter values for which the simulated summary statistics were too different from the observed ones. Estimation with smooth kernels $K$ was proposed by \citeasnoun{beaumontzhangbalding}.

%

\subsection{Regression adjustment}

Besides introducing smoothing in the ABC algorithm, \citeasnoun{beaumontzhangbalding} proposed additionally to adjust the $\theta_i$'s to weaken the effect of the discrepancy between ${\bf s}_i$ and ${\bf s}_{obs}$. In the neighborhood of ${\bf s}_{obs}$, they proposed to approximate the conditional expectation of $\theta$ given ${\bf s}$ by $\hat{m}_1$ where

\begin{equation}
\label{eq:lp}
\hat{m}_1({\bf s})=\hat{\alpha}+ ({\bf s} -{\bf s}_{obs})^t\hat{{\bf \beta}}^2\; \mbox{for} \; {\bf s} \; \mbox{such that} \; K_B({\bf s} -{\bf s}_{obs})>0.
\end{equation}

The estimates $\hat{\alpha}$ ($\alpha \in {\mathbb R}$) and $\hat{\bf \beta}$ (${\bf \beta} \in {\mathbb R}^d$) are found by minimizing the weighted sum of squared residuals

\begin{equation}
\label{eq:min}
{\rm WSSR}=\sum_{i=1}^n\{\theta_i-({\bf \alpha}+({\bf s}_i -{\bf s}_{obs})^t{\bf \beta})\} K_B({\bf s}_i -{\bf s}_{obs}).
\end{equation}

The least-squares estimate is given by \cite{rupp94}
\begin{equation}
\label{eq:sol}
(\hat{\alpha},\hat{\bf \beta})=(X^tWX)^{-1}X^tW\theta,
\end{equation}
where $W$ is a diagonal matrix whose $i^{\rm th}$ element is  $K_B({\bf s}_i -{\bf s}_{obs})$, and
$$
X=\left(
\begin{array}{cccc}
1 & s^1_1-s^1_{obs} & \cdots & s^d_1-s^d_{obs}\\
\vdots & \cdots & \ddots & \vdots \\
1 & s^1_n-s^1_{obs} & \cdots & s^d_n-s^d_{obs}
\end{array}
\right),
\;
\theta=\left(
\begin{array}{c}
\theta_1\\
\vdots\\
\theta_n
\end{array}
\right),
$$
and $s_i^j$ denotes the $j^{th}$ component of ${\bf s}_i$. The principle of regression adjustment consists of forming the empirical residuals ${\bf \epsilon}_i={\theta}_i-\hat{m}_1({\bf s}_i)$, and to adjust the $\theta_i$ by computing

\begin{equation}
\label{eq:thetastar}
\theta_i^*=\hat{m}_1({\bf s}_{obs})+\epsilon_i, \; i=1,\dots,n.
\end{equation}

Estimation of $g(\theta|{\bf s}_{obs})$ is obtained with the estimator of equation (\ref{eq:est1}) after replacing the $\theta_i$'s by the $\theta_i^*$'s. This leads to the estimator proposed by \citeasnoun[eq. (9)]{beaumontzhangbalding}

\begin{equation}
\label{eq:est2}
\hat{g}_1(\theta|{\bf s}_{obs})=\frac{\sum_{i=1}^n \tilde{K}_{b'}(\theta_i^*-\theta) K_B({\bf s}_i -{\bf s}_{obs})} {\sum_{i=1}^n K_B({\bf s}_i -{\bf s}_{obs})}.
\end{equation}

To improve the estimation of the conditional mean, we suggest a slight modification to $\hat{g}_1(\theta|{\bf s}_{obs})$ using a quadratic rather than a linear adjustment. Adjustment with general non-linear regression models was already proposed by \citeasnoun{blumfrancois} in ABC. The conditional expectation of $\theta$ given ${\bf s}$ is now approximated by $\hat{m}_2$ where

\begin{equation}
\label{eq:lp2}
 \hat{m}_2({\bf s})=\breve{\alpha}+({\bf s} -{\bf s}_{obs})^t{\breve{\beta}} + \frac{1}{2}({\bf s} -{\bf s}_{obs})^t\breve{\gamma} ({\bf s} -{\bf s}_{obs})\; \mbox{for} \; {\bf s} \; \mbox{such that} \; K_B({\bf s} -{\bf s}_{obs})>0.
\end{equation}
The three estimates $(\breve{\alpha},\breve{\beta},\breve{\gamma}) \in \mathbb{R}\times\mathbb{R}^d\times\mathbb{R}^{d^2}$ are found by minimizing the quadratic extension of the least square criterion given in (\ref{eq:min}). Because $\gamma$ is a symmetric matrix, the inference of $\gamma$  only requires the lower triangular part and the diagonal of the matrix to be estimated. The solution to this new minimization problem is given by (\ref{eq:sol}) where the design matrix $X$ is now equal to

$$
X=\left(
\begin{array}{cccccccc}
1 & s^1_1-s^1_{obs} & \cdots & s^d_1-s^d_{obs} & \frac{(s^1_1-s^1_{obs})^2}{2}  & (s^1_1-s^1_{obs})(s^2_1-s^2_{obs}) & \cdots & \frac{ (s^d_1-s^d_{obs})^2}{2}  \\
\vdots & \cdots & \ddots & \vdots & \vdots & \vdots &\ddots &\vdots  \\
1 & s^1_n-s^1_{obs} & \cdots & s^d_n-s^d_{obs} & \frac{ (s^1_n-s^1_{obs})^2}{2} & (s^1_n-s^1_{obs})(s^2_n-s^2_{obs}) & \cdots & \frac{(s^d_n-s^d_{obs})^2}{2}  \\
\end{array}
\right),
$$

Letting $\theta_i^{**}  =  \hat{m}_2({\bf s}_{obs})+(\theta_i-\hat{m}_2({\bf s}_{i}))$, the new estimator of the partial posterior distribution is given by

\begin{equation}
\label{eq:est3}
\hat{g}_2(\theta|{\bf s}_{obs})=\frac{\sum_{i=1}^n \tilde{K}_{b'}(\theta_i^{**}-\theta) K_B({\bf s}_i -{\bf s}_{obs})} {\sum_{i=1}^n K_B({\bf s}_i -{\bf s}_{obs})}.
\end{equation}
Estimators with regression adjustment in the same vein as those proposed in equations (\ref{eq:est2}) and (\ref{eq:est3}) have already been proposed  by \citeasnoun{hyndman96} and \citeasnoun{hansen04} for performing conditional density estimation when $d=1$.

\section{Asymptotic bias and variance in ABC}

\subsection{Main theorem}

To study the asymptotic bias and variance of the three estimators of the partial posterior distribution $\hat{g}_j(\cdot|{\bf s}_{obs})$, $j=0,1,2$, we assume that the bandwidth matrix is diagonal $B=b D$. A more general result for non-singular matrix $B$ is given in the Appendix. In practice, the bandwidth matrix $B$ may depend on the simulations, but we will assume in this Section that it has been fixed independently of the simulations. This assumption facilitates the computations and is classical when investigating the asymptotic bias and variance of non-parametric estimators \cite{rupp94}.

The first (resp. second) derivative of a function $f$ with respect the variable $x$ is denoted $f_x$ (resp. $f_{xx}$). When the derivative is taken with respect to a vector ${\bf x}$, $f_{\bf x}$ denotes the gradient of $f$ and $f_{\bf xx}$ denotes the Hessian of $f$.
The variance-covariance matrix of $K$ is assumed to be diagonal and equal to $\mu_2(K) I_d$. We also introduce the following notations $\mu_2(\tilde{K})=\int_u u^2\tilde{K}(u)\,du$, $R(K)=\int_{\bf u} K^2({\bf u})\,d{\bf u}$, and $R(\tilde{K})=\int_u \tilde{K}^2(u)\,du$. Finally, if $X_n$ is a sequence of random variables and $a_n$ is a deterministic sequence,  the notation $X_n=o_P(a_n)$ mean that $X_n/a_n$ converges to zero in probability and $X_n=O_P(a_n)$ means that the 
ratio $X_n/a_n$ stays bounded in the limit in probability. 
\begin{theorem}
\label{th:reject_beaumont1}
Assume that $B=b D$, in which $b>0$ is the bandwidth associated to the kernel $K$, and assume that conditions (A1):(A5) of the Appendix hold. The bias and the variance of the estimators $\hat{g}_j(\cdot|{\bf s}_{obs})$, $j=0,1,2$, are given by
\begin{equation}
\label{eqn:bias_rej_beaumont}
E[\hat{g}_j(\theta|{\bf s}_{obs})-g(\theta|{\bf s}_{obs})]  =  C_1{b'}^2+C_{2,j}b^2 + O_P((b^2+b'^2)^2)+O_P(\frac{1}{n|B|}),  \; j=0,1,2,
\end{equation}

\begin{equation}
\label{eqn:var_rej_beaumont}
{\rm Var}[\hat{g}_j(\theta|{\bf s}_{obs})]=\frac{C_3}{nb^db'} (1+o_P(1)), \; j=0,1,2,
\end{equation}

with
$$
C_1=\frac{\mu_2(\tilde{K})g_{\theta \theta}(\theta|{\bf s}_{obs})}{2},
 $$
\begin{equation}
\label{eq:c20}
C_{2,0}=\mu_2(K)\left(\frac{g_{\bf s}(\theta|{\bf s})^t_{|{\bf s}={\bf s}_{obs}}D^2 p_{\bf s}({\bf s}_{obs}) } {p({\bf s}_{obs})}+\frac{{\rm tr}(D^2 g_{\bf ss}(\theta|{\bf s})_{|{\bf s}={\bf s}_{obs}})} {2}\right), 
\end{equation}
\begin{equation}
\label{eq:c21}
C_{2,1}=\mu_2(K) \left(
\frac{h_{\bf s}(\epsilon|{\bf s})_{|{\bf s}={\bf s}_{obs}}^t D^2 p_{\bf s}({\bf s}_{obs}) } {p({\bf s}_{obs})}+
\frac{{\rm tr}(D^2 h_{\bf ss}(\epsilon|{\bf s})_{|{\bf s}={\bf s}_{obs}})}{2}-
\frac{h_{\epsilon}(\epsilon|{\bf s}_{obs}) {\rm tr}(D^2 m_{\bf ss}({\bf s}_{obs})) } {2}
\right),
\end{equation}

\begin{equation}
\label{eq:c22}
C_{2,2}=\mu_2(K) \left(
\frac{h_{\bf s}(\epsilon|{\bf s})^t_{|{\bf s}={\bf s}_{obs}}D^2 p_{\bf s}({\bf s}_{obs}) }{p({\bf s}_{obs})}+
\frac{{\rm tr}(D^2 h_{\bf ss}(\epsilon|{\bf s})_{|{\bf s}={\bf s}_{obs}})}{2}
\right),
\end{equation}
and
\begin{equation}
\label{eqn:c3}
C_3=\frac{R(K) R(\tilde{K}) g(\theta|{\bf s}_{obs})}{|D|p({\bf s}_{obs})}.
\end{equation}
\end{theorem}

{\par{\bf Remark 1. Curse of dimensionality}}
The mean square error (MSE) of an estimator is equal to the sum of its squared bias and its variance. With standard algebra, we find that the MSEs of the three estimators $\hat{g}_j(\cdot|{\bf s}_{obs})$, $j=0,1,2$, are minimized when both $b$ and $b'$ are of the order of $n^{-1/(d+5)}$. This implies that the minimal MSEs are of the order of $n^{-4/(d+5)}$. Thus, the rate at which the minimal MSEs converge to 0 decreases importantly as the dimension $d$ of ${\bf s}_{obs}$ increases. However, we wish to add words of caution since this standard asymptotic argument does not account for the fact that the `constants' $C_1,C_2,C_3$ involved in Theorem \ref{th:reject_beaumont1} also depend on the dimension of the summary statistics. Moreover, in the context of multivariate density estimation, \citeasnoun{Scott92} argued that conclusions arising from the same kind of theoretical arguments were in fact much more pessimistic than the empirical evidence. Finally, because the underlying structure of the summary statistics can typically be of dimension lower than $d$, dimension reduction techniques, such as neural networks or partial least squares regression, have been proposed \cite{blumfrancois,Wegmannetal}.

{\par{\bf Remark 2. Effective local size and effect of design}} As shown by equations (\ref{eqn:var_rej_beaumont}) and (\ref{eqn:c3}), the variance of the estimators can be expressed, up to a constant, as $\frac{1}{\tilde{n}}\frac{g(\theta|{\bf s}_{obs})}{b'}$, where the effective local size is $\tilde{n}=n|D|p({\bf s}_{obs})b^d$. The effective local size is an approximation of the expected number of simulations that fall within the ellipsoid of radii equal to the diagonal elements of $D$ times $b$. Thus equations (\ref{eqn:var_rej_beaumont}) and (\ref{eqn:c3}) reflect that the variance is penalized by sparser simulations around ${\bf s}_{obs}$. Sequential Monte Carlo samplers \cite{sissonfantanaka,robertbeaumontmarincornuet,tonietal} precisely aim at adapting the sampling distribution of the parameters, a.k.a. the design, to increase the probability of targeting close to ${\bf s}_{obs}$. Likelihood-free MCMC samplers have also been proposed to increase the probability of targeting close to ${\bf s}_{obs}$  \cite{marjorametal03,sissonfan}.

{\par{\bf Remark 3. A closer look at the bias}}
There are two terms in the bias of $\hat{g}_0(\cdot|{\bf s}_{obs})$ (equation (\ref{eq:c20})) that are related to the smoothing in the space of the summary statistics. The first term in equation (\ref{eq:c20}) corresponds to the effect of the design and is large when the gradient of $Dp(\cdot)$ is collinear to the gradient of $Dg(\theta|\cdot)$. This term reflects that, in the neighborhood of ${\bf s}_{obs}$, there will be an excess of points in the direction of $Dp_{\bf s}({\bf s}_{obs})$. Up to a constant, the second term in equation (\ref{eq:c20}) is proportional to ${\rm tr}(D^2 g_{\bf ss}(\theta|{\bf s})_{|{\bf s}={\bf s}_{obs}})$ which is simply the sum of the elementwise product of $D$ and the Hessian $g_{\bf ss}(\theta|{\bf s})_{|{\bf s}={\bf s}_{obs}}$. This second term shows that the bias is increased when there is more curvature of $g(\cdot |{\bf s})$ at ${\bf s}_{obs}$ and more smoothing.

For the estimator $\hat{g}_2(\cdot|{\bf s}_{obs})$ with quadratic adjustment, the asymptotic bias is the same as the bias of an estimator for which the conditional mean would be known exactly. Results of the same nature were found, for $d=1$, by \citeasnoun{fan98} when estimating the conditional variance and by \citeasnoun{hansen04} when estimating the conditional density. Compared to the bias of $\hat{g}_2(\cdot|{\bf s}_{obs})$, the bias of the estimator with linear adjustment $\hat{g}_1(\cdot|{\bf s}_{obs})$ contains an additional term depending on the curvature of the conditional mean.

\subsection{Bias comparison between the estimators with and without adjustment}

To investigate the differences between the three estimators, we first assume that the partial posterior distribution of $\theta$ can be written as $h(\theta-m({\bf s}))$ in which the function $h$ does not depend on ${\bf s}$. This amounts to assuming an homoscedastic model in which the conditional distribution of $\theta$ given ${\bf s}$ depends on ${\bf s}$  only through the conditional mean $m({\bf s})$. If the conditional mean $m$ is linear in ${\bf s}$, both $C_{2,1}$ and $C_{2,2}$ are null involving that both estimators with regression adjustment have a smaller bias than $\hat{g}_0(\cdot|{\bf s}_{obs})$. For such ideal models, the bandwidth $b$ of the estimators with regression adjustment can be taken infinitely large so that the variance will be inversely proportional to the total number of simulations $n$. Still assuming that $g(\theta|{\bf s})=h(\theta-m({\bf s}))$, but with a non-linear $m$, the constant $C_{2,2}$ is null so that the estimator $\hat{g}_2(\cdot|{\bf s}_{obs})$ has the smallest asymptotic MSE. However, for general partial posterior distributions, it is not possible to rank the three different biases. Consequently, when using the estimators with adjustment, the parameterization of the model should be guided toward making the distributions $g(\theta|{\bf s})$ as homoscedastic as possible. To achieve this objective, we propose, in the next section, to use transformations of the summary statistics.

\section{Choosing a regression model}

\subsection{Transformations of the summary statistics and the parameters}

To make the regression as homoscedastic as possible, we propose to transform the summary statistics in equations ($\ref{eq:lp}$) and ($\ref{eq:lp2}$). Here we consider logarithmic and square root transformations only. We choose the transformations that minimize the weighted sum of squared residuals (WSSR) given in equation (\ref{eq:min}) in which we take an uniform kernel for the weight function $K$. The weights $K_B({\bf s}_i -{\bf s}_{obs})$ depend on the transformations of the summary statistics and the uniform kernel ensures that the WSSR are comparable for different transformations. Since there are a total of $3^d$ regression models to consider, greedy algorithm can be considered for large values of $3^d$. 

Although transformations of the parameter $\theta$ in the regression equations ($\ref{eq:lp}$) and ($\ref{eq:lp2}$) can also stabilize the variance \cite{boxcox64}, we rather use transformations of $\theta$ for guaranteeing that the adjusted parameters $\theta_i^*$ and $\theta_i^{**}$ lie in the support of the prior distribution \cite{beaumontzhangbalding}. For positive parameters, we use a $\log$ transformation before regression adjustment. After adjusting the logarithm of a positive parameter, we return to the original scale using an exponential transformation. Replacing the logarithm by a logit transformation, we consider the same procedure for the parameters for which the support of the prior is a finite interval. 

\subsection{Choosing an estimator of $g(\cdot | {\bf s}_{obs})$}

In Section 3, we find that there is not a ranking of the three estimators $\hat{g}_j(\cdot|{\bf s}_{obs})$, $j=0,1,2$, which is universally valid.  Since the three estimators rely on local regressions, of degree 0, 1, and 2, we propose to choose the regression model that minimizes the prediction error of the regression. Because the regression models involve a different number of predictors, we use cross-validation to evaluate the prediction error. We introduce the following leave-one-out estimate
$$
CV_j=\sum_{i=1}^n (\hat{m}_j^{-i}({\bf s}_{i})-\theta_i),  \; j=0,1,2,
$$
where $\hat{m}_j^{-i}({\bf s}_{i})$ denotes the estimate of $m(\theta_i | {\bf s}_{i})$ obtained, in the neighborhood of ${\bf s}_{i}$, with a local polynomial of degree $j$ by removing the $i^{th}$ point of the training set.

\section{Examples}

\subsection{Example 1:  A Gaussian model}
We are interested here in the estimation of the variance parameter $\sigma^2$ in a Gaussian sample. Although Approximate Bayesian Computation is not required for such a simple model, this example will highlight the potential importance of the transformations of the summary statistics and of the methods with adjustment.  Assume that we observe a sample of size $N=50$ in which each individual is a Gaussian random variable of mean $\mu$ and variance $\sigma^2$. We assume the following hierarchical prior for $\mu$ and $\sigma^2$ \cite{Gelman03}
\begin{eqnarray}
\sigma^2 & \sim & {\rm Inv}\chi^2({\rm d.f.}=1)\\
\mu & \sim & \mathcal{N}(0,\sigma^2),
\end{eqnarray}
where ${\rm Inv}\chi^2({\rm d.f.}=\nu)$ denotes the inverse chi-square distribution with $\nu$ degrees of freedom, and $\mathcal{N}$ denotes the Gaussian distribution. We consider the empirical mean $\bar{x}_N$ and variance $s_N^2$ as the summary statistics. These two statistics are sufficient with respect to the parameter $\sigma^2$ \cite{Gelman03}. The data come from the well-known Iris data set and consist of the sample of the petal lengths for the virginica species ($\bar{x}_N=5.552$, $s_N^2=0.304$, and $N=50$).

\begin{figure}[h]
\begin{center}
\includegraphics[height = 7.5cm]{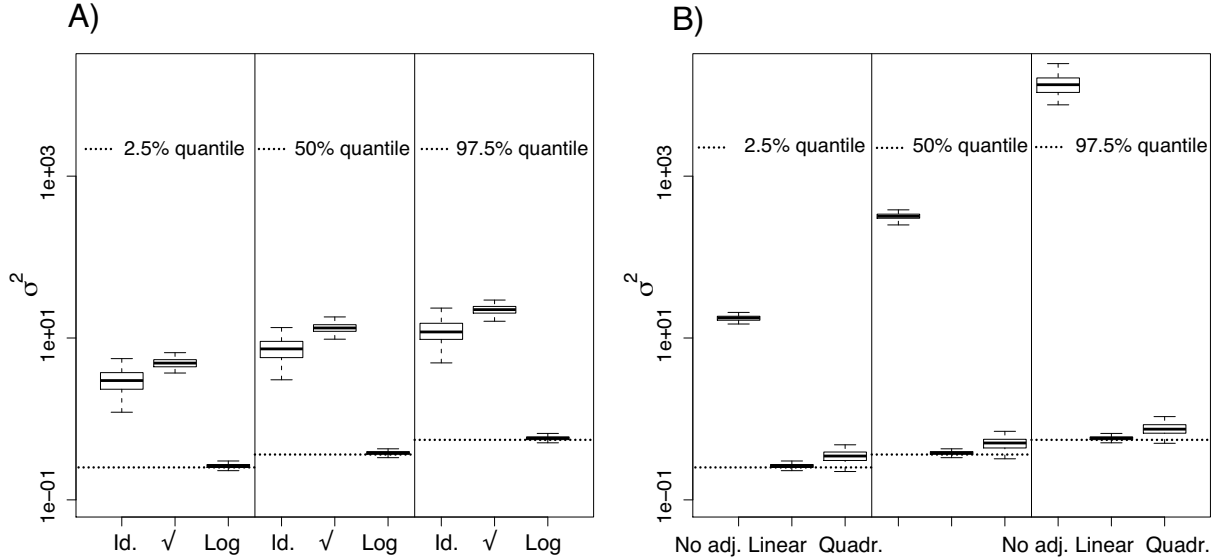}
\end{center}
\caption{Estimation of the posterior quantiles of  the variance parameter $\sigma^2$ in a Gaussian sample. We perform a total of 100 ABC replicates and we display the boxplots of the estimated posterior quantiles. A)Estimation of the posterior quantiles with linear adjustment using $(\bar{x}_N,s_N^2)$, $(\bar{x}_N,\sqrt{s_N^2})$, and $(\bar{x}_N,\log{s_N^2})$. B) Estimation of the posterior quantiles with no adjustment and with linear and quadratic adjustment  considering $(\bar{x}_N,\log{s_N^2})$ as the summary statistics. The horizontal lines correspond to the true posterior quantiles. In this Gaussian example, both $\log$ transformation of the empirical variance and regression adjustment are crucial for accurate estimation of the posterior distribution. Id. stands for the identity function, adj. for adjustment and quadr. for quadratic.} 
\label{fig:fig1}
\end{figure}

We perform a total of 100 ABC replicates. Each replicate consists of simulating $n=20,000$ Gaussian samples. We consider a spherically symmetric kernel for $K$ and an Epanechnikov kernel for $K_1$. We assume a diagonal bandwidth matrix $B=b D$ where $D$ contains the standard deviation of each summary statistic in the diagonal and $b$ is the $2.5\%$ quantile of the distances $\|{\bf s}_i -{\bf s}_{obs}\|$, $i=1 \cdots n$. This procedure amounts to choosing the 500 simulation that provide the best match to the observed summary statistics. In the two following examples, we consider the same number of simulations, the same bandwidth matrix, and the same kernel. Here the true posterior distribution is known exactly \cite{Gelman03} and can be compared to the different estimates obtained with ABC. Since $\sigma^2$ is a positive parameter, its $\log$ is regressed as described in Section 4.   
As displayed by Figure \ref{fig:fig1}, the estimate with linear adjustment $\hat{g}_1(\sigma^2| \bar{x}_N,  s_N^2)$ provides a good estimate provided that the empirical variance is log-transformed in the regression setting. The WSSR criterion selects the right transformation here since it is minimum for the logarithmic transformation in all of the 100 test replicates. When considering $\bar{x}_N$ and $\log{s_N^2}$ in the regression, both the linear and the quadratic adjustment provide good estimate of $\sigma^2$ by contrast to the method without adjustment (see Figure \ref{fig:fig1}). The cross-validation criterion never selects the method without adjustment, selects 74 times linear adjustment and 26 times quadratic adjustment. 

\subsection{Example 2:  Coalescent model in population genetics}

\begin{figure}[h]
\begin{center}
\includegraphics[height = 7cm]{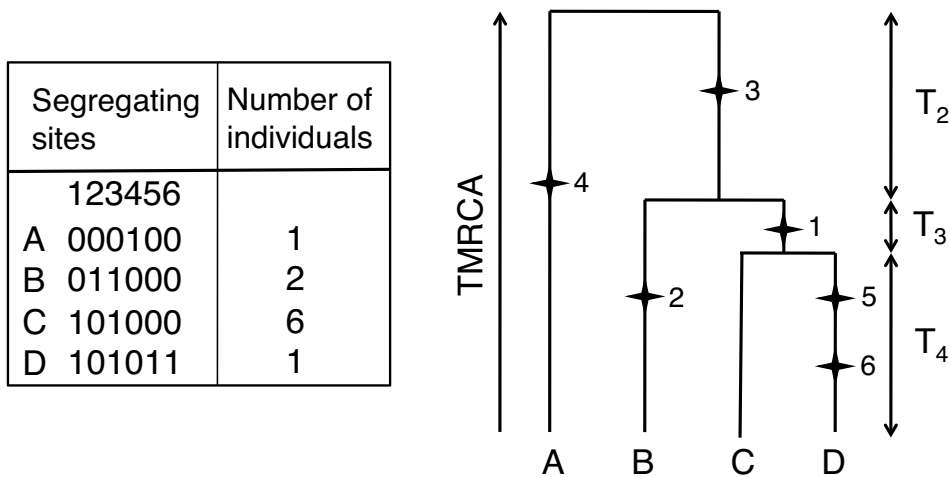}
\end{center}
\caption{Coalescent process for simulating DNA sequences. This example is excerpted from Cox (2008). There are a total of ten DNA sequences. We display only  the upper part of the tree in which mutations occur. We omit the lower part corresponding to the coalescence times $T_5,\dots ,T_{10}$. The ancestral sequence is a sequence of 500 base pairs and contains a repetition of 0. The stars denote the $0 \rightarrow 1$ mutations. To generate this tree, a mutation rate of $3.6 \times 10^{-6}$/base pairs/generation was considered. The true TMRCA is equal to $465$ generations here. To infer the TMRCA, we consider the number of segregating sites $S=6$ and the mean number of mutations between the ancestor and the individuals $\rho = 2.10$ as summary statistics.} 
\label{fig:tmrca}
\end{figure}

ABC was originally developed for inferring parameters of coalescent models in populations genetics \cite{pritchardetal}. Coalescent models describe, in a probabilistic fashion, the tree-like ancestry of genes represented in a sample. Because the ancestral tree is unknown, the likelihood involves an integral over this high dimensional ancestral tree and is computationally intractable. Here we aim at estimating the Time since the Most Recent Common Ancestor (TMRCA) of a sample of gene. This time is equal to the age of the root of the ancestral tree. A description of the coalescent process and the TMRCA is given in Figure \ref{fig:tmrca}. The coalescent prior for the TMRCA and the whole ancestral tree can be described by the following hierarchical procedure 
\begin{enumerate}
\item Simulate the size of the entire population $N$ according to its prior distribution, a uniform distribution between 0 and 10,000 here.
\item Simulate the $T_k$'s, the $k^{th}$ inter-coalescence times, as exponential random variables of rate $k(k-1)/(2N), k = 2 \dots m$, where $m$ is the number of sequences in the sample. Time is counted in generations here.
\end{enumerate}
The TMRCA is given by the sum of the inter-coalescence times $T_2+\dots +T_m$. Once the genealogical tree has been generated, DNA sequences are simulated by superimposing mutations along the tree according to a Poisson process of rate $u$ where $u$ is the mutation rate. Here we assume that the mutation rate is known and we use $u=1.8 \times 10^{-3}$ mutation/generation for the whole 500 base pairs DNA sequence \cite{cox08}. Assuming the {\it infinitely-many-sites} model, each mutation hit a so-called segregating site that has never been hit before. As summary statistics, we consider the total number of segregating sites $S$ and the mean number of mutations between the ancestor and each individual in the sample. The latter summary statistic is called the $\rho$ statistic and is central in the field of molecular dating \cite{cox08}.

\begin{figure}[h]
\begin{center}
\includegraphics[height = 7cm]{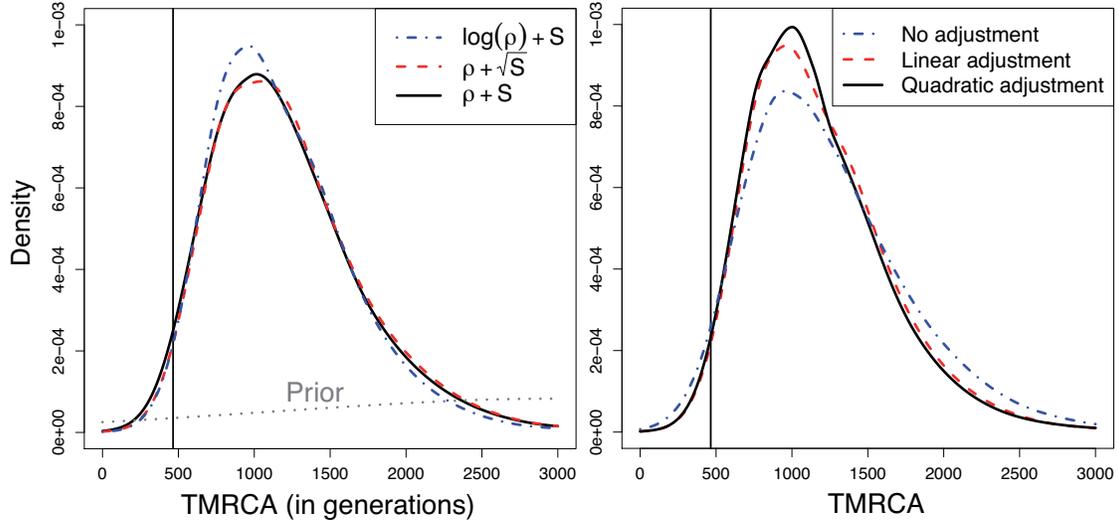}
\end{center}
\caption{Posterior distribution of the TMRCA. A) Estimated posterior distributions with linear adjustment considering three different transformations of the summary statistics. The summary statistics $\log{\rho}$ and $S$ provide the smallest residual error. B) Estimated posterior distributions using the three different estimates $\hat{g}_j({\rm TMRCA}|(\log{\rho},S))$, $j=0,1,2$. The quadratic regression provides the smallest prediction error as found with a leave-one-out estimate. For this coalescent example, both transformations of the summary statistics and regression adjustments do not greatly alter the estimated posterior distribution}
\label{fig:post2}
\end{figure}

We infer the TMRCA using the DNA sequences simulated by  \citeasnoun{cox08}. The true TMRCA was equal to $465$ generations in his simulation and the values of the summary statistics are $S=6$ and $\rho = 2.10$. Since the TMRCA is a positive parameter, we use a logarithmic transformation when performing the regression adjustment. As shown in Table $\ref{tab:Tab1}$, the WSSR criterion selects the regression equation $\log{\rm TMRCA}=\log{\rho}+S$. The cross validation criterion points to the estimator with quadratic adjustment although the prediction errors obtained with the linear and quadratic regressions are almost the same (see Table $\ref{tab:Tab2}$). In this example, we do not observe the dramatic effect of the transformations and the adjustments that we found for the Gaussian example.  As displayed in Figure \ref{fig:post2}, both transformations of the summary statistics and regression adjustments do not greatly alter the estimated posterior distribution. Figure \ref{fig:post2} also shows that the posterior distribution is clearly more peaky than the prior indicating that the summary statistics convey substantial information about the TMRCA. The $95\%$ credibility interval of the posterior  $(400-2450)$ is indeed considerably narrower than the credibility interval of the prior $(300-30,800)$. However, as is typical with molecular dating, there remains considerable uncertainty when estimating the TMRCA  \cite{cox08}. The $95\%$ credibility interval of the TMRCA ranges from a value slightly inferior to the true one to a value more than five times larger than the true one.

\begin{table}[h]
\begin{center}
\caption{Choosing a transformation of the summary statistics}
\fontsize{6.5}{7}\selectfont
\label{tab:Tab1}

\begin{tabular}{c|ccccccccc}
   
   {\small Parameter} &\multicolumn{9}{c}{\small Sum of squared residuals} \\
   \hline  
 TMRCA & $\rho +S$ & $\rho +\sqrt{S}$ & $\rho +\log{S}$ &  $\sqrt{\rho} +S$ &  $\sqrt{\rho} +\sqrt{S}$   & $ \sqrt{\rho} +\log{S}$ & $ \mbf{\log{\rho} +S}$ & \textbf{$\log{\rho} +\sqrt{S}$} & $\log{\rho} +\log{S}$\\  
 Example 2 & 0.19&0.19&0.18& 0.18&0.18& 0.18& \textbf{0.16}& 0.17 & 0.17 \\  
   \hline
  
  Transmission rate $\alpha-\delta$ & $G+H$ & $G +\sqrt{H}$ & $G +\log{H}$ &  $\sqrt{G} +H$ &  $\sqrt{G} +\sqrt{H}$   & $ \sqrt{G} +\log{H}$ & $ \log{G} +H$ & \textbf{$\log{G} +\sqrt{H}$} & $\mbf{\log{G} +\log{H}}$\\  
Example 3&0.48 & 0.48 & 0.15 &  0.47 & 0.48 &  0.15 & 0.47 & 0.48 &   \textbf{0.14}\\
     \hline
     
$R_0=\alpha/\delta$& $G +H$ & $G +\sqrt{H}$ & $\mbf{G +\log{H}}$ &  $\sqrt{G} +H$ &  $\sqrt{G} +\sqrt{H}$   & $ \sqrt{G} +\log{H}$ & $ \log{G} +H$ & \textbf{$\log{G} +\sqrt{H}$} & $\log{G} +\log{H}$\\
 Example 3&1.53 & 1.54 & \textbf{1.33} & 1.51 & 1.49 &  1.34 &  1.51 & 1.47 & 1.34\\
   \hline
 
\end{tabular}
\end{center}
\end{table}

\begin{table}[h]
\begin{center}
\caption{Cross validation criterion for choosing an estimator of the posterior distribution}
\label{tab:Tab2}

\begin{tabular}{c|ccc}
Parameter   & No adjustment &Linear adjustment & Quadratic adjustment\\
  \hline

 TMRCA (Example 2) &   0.90 &0.624 &\textbf{0.620}\\  
  
Transmission rate $\alpha-\delta$ (Example 3) &   1.92 &  \textbf{0.31} & 0.34\\  
$R_0=\alpha/\delta$ (Example 3)&  2.15 & \textbf{1.53} &  1.65\\  
\hline

\end{tabular}
\end{center}
\end{table}

\subsection{Example 3:   Birth and death process in epidemiology}

To study the rate at which tuberculosis spread in a human population, \citeasnoun{tanakafrancislucianisisson} make use  of available genetic data  of {\it Mycobacterium tuberculosis} isolated from different patients.  DNA fingerprint at the {\it IS6110} marker were obtained for 473 isolates sampled in San Francisco during 1991 and 1992 \cite{smalletal}.  The {\it IS6110} fingerprints were grouped into 326 distinct genotypes whose configuration into clusters is represented by
$$
30^1 23^1 15^1 10^1 8^1 5^2 4^4 3^{13} 2^{20} 1^{282},
$$
where $n^k$ indicates there are $k$ clusters of size $n$. To infer the rate of transmission of the disease from this data, \citeasnoun{tanakafrancislucianisisson} introduced a stochastic model of transmission and mutation. We denote by $X_i(t)$ the number of cases of type $i$ at time $t$, by $G(t)$ the current number of distinct genotypes, and by $N(t)$ the total number of cases. The  model starts with $X_1(0)=1$, $N(0) = 1$ and $G(0) = 1$. We denote by $\alpha$, $\delta$, and $\theta$, the per-capita birth rate, death rate and mutation rate. When a birth occurs for an individual of genotype $i$, the value of $X_i(t)$ is incremented by 1. If the event is a death,  the value of $X_i(t)$ is decremented by 1. When a mutation occurs for an individual of genotype $i$, we assume the {\it infinitely-many-alleles} model in which a new allele is formed. This means that the value of  $X_i(t)$ is decremented by 1 and a case of a new genotype is created. Following \citeasnoun{tanakafrancislucianisisson}, the process is stopped when $N=10,000$. At the stopping time,  a sample of size $n=473$ is drawn from the final population randomly without replacement. As summary statistics, we consider the total number of genotypes $G$ in the sample and the {\it homozygosity} $H$ of the sample defined as $H=\sum (n_i/n)^2$, where $n_i$, $i=1 \dots  G,$ denotes the number of individual of genotype $i$ in the sample. We consider the following prior specification

\begin{eqnarray}
& &\theta  \sim  \mathcal{N}(0.20,0.07^2) \\
&&(\frac{\alpha}{\alpha+\delta+\theta},\frac{\delta}{\alpha+\delta+\theta},\frac{\theta}{\alpha+\delta+\theta})  \sim  {\rm Dir}(1,1,1) \,  | \, \delta<\alpha.
\end{eqnarray}
The informative prior for $\theta$ (in mutation/year) arises from previous estimations of the mutation rate \cite{tanakafrancislucianisisson}.

\begin{figure}[h]
\begin{center}
\includegraphics[height = 5cm]{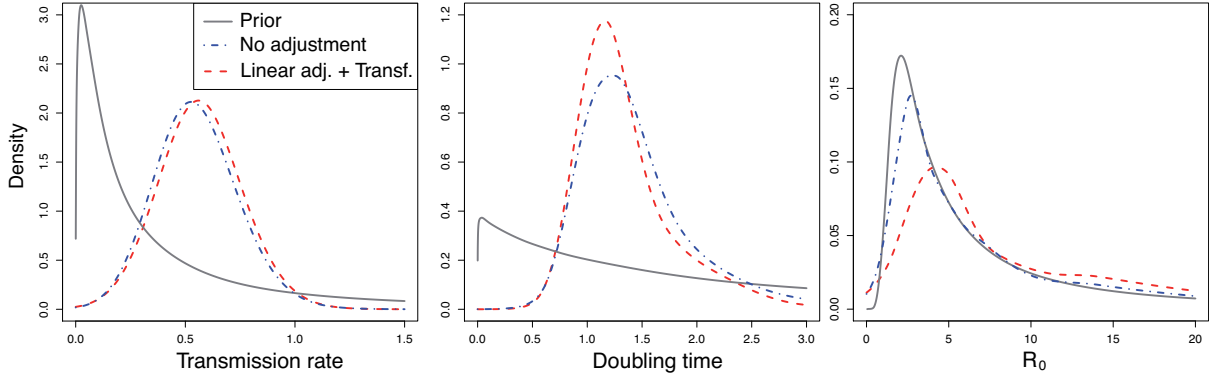}
\end{center}
\caption{Posterior distributions of key epidemiological quantities for the tuberculosis epidemic in San Francisco. The abbreviation transf. stands for transformation. In this example, both transformations of the summary statistics and regression adjustments do not greatly alter the estimated posterior distributions except when estimating $R_0$. For the transmission rate and the doubling time, the posterior distributions greatly differ from the prior distributions. For the reproduction number $R_0$, there is not an important difference between the prior and the posterior indicating than the data do not convey enough information for a confident estimation of $R_0$.}
\label{fig:post3}
\end{figure}

We are interested in the estimation of the net transmission rate $\alpha - \delta$, of the doubling time of the disease $\log{2}/(\alpha - \delta)$, and of the basic reproduction number $R_0=\alpha / \delta$. Since they are positive parameters, they are $\log$-transformed in the regression equations. Once $\log$-transformed, the transmission rate and the doubling time are equal up to a multiplicative constant so that the optimal transformation and adjustment are the same for both parameters. We find that transforming $G$ and $H$ with the $\log$ function is optimal for inferring the doubling time whereas $\log$-transforming $H$ only is optimal for inferring $R_0$ (see Table \ref{tab:Tab1}). For all parameters, we select linear adjustment based on the cross-validation criterion (see Table \ref{tab:Tab2}).  As displayed in Figure \ref{fig:post3}, transformations of the summary statistics and regression adjustments do not greatly alter the estimated posterior distributions except when estimating $R_0$. For the transmission rate and the doubling time, the posterior distributions greatly differ from the prior distributions (see Figure  \ref{fig:post3} and Table \ref{tab:Tab3}).  However, for the reproduction number $R_0$, the posterior $95\%$ credibility interval is hardly narrower than the prior credibility interval. These comparisons between the prior and the posterior distributions suggest that the genotype data convey much more information for estimating the transmission rate and the doubling time than for estimating the reproduction number $R_0$. A large credibility interval for the parameter $R_0$ was also find by \citeasnoun{tanakafrancislucianisisson}.

\begin{table}[h]
\begin{center}
\caption{Posterior estimates of epidemiological quantities for the San Francisco data} 
\label{tab:Tab3}
\begin{tabular}{c|c|cccccccc}
Parameter & Description & $95\%$ Prior C.I.$^{a}$ & Posterior mode & $95\%$ Posterior C.I.$^{a}$ \\
\hline
$\alpha-\delta$ & Transmission rate (years) & 0.01-9.97 & 0.56 & 0.16-0.95  \\
$\log{2}/(\alpha-\delta)$ & Doubling time (years) & 0.06-57.85 & 1.16 & 0.73-4.35  \\
$\alpha/\delta$ & Reproduction number $R_0$ & 1.27-123.32 & 4.00 & 2.24-117.45\\
\hline
\end{tabular}
\end{center}
$^{a}$ C.I. stands for credibility intervals 
\end{table}

\section{Conclusion}

In this paper, we presented Approximate Bayesian Computation as a technique of inference that relies on stochastic simulations and non-parametric statistics. We introduced an estimator of $g(\theta|{\bf s}_{obs})$ based on quadratic adjustment for which the asymptotic bias involves fewer terms than the asymptotic bias of the estimator with linear adjustment proposed by Beaumont et al. (2002). More generally, we showed that the benefit arising from regression adjustment (equations (\ref{eq:est2}) and (\ref{eq:est3})) is all the more important that the distribution of the residual $\epsilon$ is independent of ${\bf s}$ in the regression model $\theta({\bf s})=m({\bf s})+\epsilon$. 
To make this regression model as  homoscedastic as possible, we proposed to use transformations of the summary statistics when performing regression adjustment. We proposed to select the transformation of the  summary statistics that minimizes the sum of squared residuals within the window of the accepted simulations. 
In a Gaussian example, we showed that transformations of the summary statistics and regression adjustment can dramatically improve inference in ABC. In two other examples borrowed from the population genetics and epidemiology literature, regression adjustment and transformations of the summary statistics had little effect on the estimated posterior distribution. However, above all, these two examples emphasize the potential of ABC for complex models for which the likelihood is not computationally tractable.

\appendix

\makeatletter
\begin{center}
APPENDIX
\end{center}
 \renewcommand{\@seccntformat}[1]{APPENDIX~{\csname the#1\endcsname}.\hspace*{1em}}
 \makeatother

\section{Hypotheses of Theorem 1}

\begin{itemize}
\item[{\bf A1)}] The kernel $K$ has a finite second order moment such that $\int{\bf uu}^TK({\bf u})\,d{\bf u}=\mu_2(K)I_d$ where  $\mu_2(K) \neq 0$. We also require that all first-order moments of $K$ vanish, that is, $\int{\bf u}_iK({\bf u})\,d{\bf u}=0$ for $i=1,\dots,d$. As noted by \citeasnoun{rupp94}, this condition is fulfilled by spherically symmetric kernels and product kernels based on symmetric univariate kernels.
\item[{\bf A2)}] The kernel $\tilde{K}$ is a symmetric univariate kernel with finite second order moment $\mu_2(\tilde{K})$.
\item[{\bf A3)}]
The observed summary statistics ${\bf s}_{obs}$ lie in the interior of the support of $p$.  At ${\bf s}_{obs}$, all the second order derivatives of the function $p$ exist and are continuous.
\item[{\bf A4)}] The point $\theta$ is in the support of the partial posterior distribution. At the point $(\theta,{\bf s}_{obs})$, all the second order derivatives of the partial posterior $g$ exist and are continuous. The conditional mean of $\theta$, $m({\bf s})$, exists in a neighborhood of ${\bf s}_{obs}$ and is finite. All its second order derivatives exist and are continuous.

\item[{\bf A5)}] The sequence of non-singular bandwidth matrices $B$ and bandwidths $b'$ is such that $1/(n|B|b')$, each entry of $B^tB$, and $b'$ tend to $0$ as $n -> \infty$.

\end{itemize}

\section{Proof of Theorem 1}

The three estimators of the partial posterior distribution $\hat{g}_j(\cdot|{\bf s}_{obs})$, $j=0,1,2$,  are all of the Nadaraya-Watson type. The difficulty in the computation of the bias and the variance of the Nadaraya-Watson estimator comes form the fact that it is a ratio of two random variables. Following \citeasnoun[p. 98]{pagan99} or \citeasnoun{Scott92}, we linearize the estimators in order to compute their biases and their variances. We write the estimators of the partial posterior distribution $\hat{g}_j$, $j=0,1,2$,  as
$$
\hat{g_j}(\theta|{\bf s}_{obs})=\frac{\hat{g}_{j,{\rm N}}} {\hat{g}_{{\rm D}}},\quad j=0,1,2,
$$
where
$$
\hat{g}_{0,{\rm N}}=\frac{1}{n}\sum_{i=1}^n \tilde{K}_{b'}(\theta_i-\theta) K_B({\bf s}_i -{\bf s}_{obs}),
$$

$$
\hat{g}_{1,{\rm N}}=\frac{1}{n}\sum_{i=1}^n \tilde{K}_{b'}(\theta_i^*-\theta) K_B({\bf s}_i -{\bf s}_{obs}),
$$

$$
\hat{g}_{2,{\rm N}}=\frac{1}{n} \sum_{i=1}^n \tilde{K}_{b'}(\theta_i^{**}-\theta) K_B({\bf s}_i -{\bf s}_{obs}),
$$
and
$$
\hat{g}_{\rm D}=\sum_{i=1}^n K_B({\bf s}_i -{\bf s}_{obs}).
$$
To compute the asymptotic expansions of the moments of the three estimators, we use the following lemma
\begin{lemme}
\label{lem:linearize}
For $j=0,1,2$, we have
\begin{eqnarray}
\label{eq:linearize}
\hat{g_j}(\theta|{\bf s}_{obs}) & = & \frac{E[\hat{g}_{j,{\rm N}}]}{E[\hat{g}_{\rm D}]}+
\frac{\hat{g}_{j,{\rm N}}-E[\hat{g}_{j,{\rm N}}]}{E[\hat{g}_{\rm D}]}-
\frac{E[\hat{g}_{j,{\rm N}}](\hat{g}_{\rm D}-E[\hat{g}_{\rm D}])}{E[\hat{g}_{\rm D}]^2} \nonumber \\
& & +
O_P({\rm Cov}(\hat{g}_{j,{\rm N}},\hat{g}_{\rm D}) +{\rm Var}[\hat{g}_{\rm D}])
\end{eqnarray}

\end{lemme}

\proof
Lemma \ref{lem:linearize} is a simple consequence of the Taylor expansion for the function $(x,y)->x/y$ in the neighborhood of the point $(E[\hat{g}_{j,{\rm N}}],E[\hat{g}_{\rm D}])$ (see \citeasnoun{pagan99} for another proof). The order of the reminder follows from the weak law of large numbers.

\eop

The following Lemma gives an asymptotic expansion of all the expressions involved in equation (\ref{eq:linearize}).
\begin{lemme}
\label{lem:list}
Suppose assumption (A1)-(A5) hold, denote $\epsilon=\theta-m({\bf s}_{obs})$, then we have
\begin{equation}
\label{eq:list1}
 E[\hat{g}_{\rm D}]  =  p({\bf s}_{obs})+\frac{1}{2}\mu_2(K){\rm tr}(BB^t p_{ss}({\bf s}_{obs})) + o({\rm tr}(B^tB)),
\end{equation}
\begin{eqnarray}
\label{eq:list2}
 & E[\hat{g}_{0,{\rm N}}]   =  p({\bf s}_{obs})g(\theta|{\bf s}_{obs}) +\frac{1}{2}b'^2 \mu_2(\tilde{K}) g_{\bf \theta\theta}(\theta|{\bf s}_{obs}) p({\bf s}_{obs}) \nonumber \\
& +  \mu_2(K)[g_{\bf s}(\theta|{\bf s})^t_{|{\bf s}={\bf s}_{obs}} BB^t p_{\bf s}({\bf s}_{obs})  +\frac{1}{2}g(\theta|{\bf s}_{obs}) {\rm tr}(BB^t p_{\bf ss}({\bf s}_{obs}))  \nonumber\\
&  + \frac{1}{2}p({\bf s}_{obs}){\rm tr}(BB^t g_{\bf ss}(\theta|{\bf s})_{|{\bf s}={\bf s}_{obs}}) ] +o(b'^2)+o({\rm tr}(B^tB)),
\end{eqnarray}
\begin{eqnarray}
\label{eq:list3}
 & E[\hat{g}_{1,{\rm N}}]  =  p({\bf s}_{obs})h(\epsilon|{\bf s}_{obs}) +\frac{1}{2}b'^2 \mu_2(\tilde{K}) h_{\epsilon\epsilon}(\epsilon|{\bf s}_{obs}) p({\bf s}_{obs}) \nonumber \\
 & + \mu_2(K)[h_{\bf s}(\epsilon|{\bf s})^t_{|{\bf s}={\bf s}_{obs}} BB^t p_{\bf s}({\bf s}_{obs})  +\frac{1}{2}h(\epsilon|{\bf s}_{obs}){\rm tr}(BB^t p_{\bf ss}({\bf s}_{obs}))  \nonumber\\
 & + \frac{1}{2}p({\bf s}_{obs}){\rm tr}(BB^t h_{\bf ss}(\epsilon|s)_{|{\bf s}={\bf s}_{obs}}) - \frac{h_{\epsilon}(\epsilon|{\bf s}_{obs})} {2} {\rm tr}(BB^t m_{\bf ss}({\bf s}_{obs}))]\nonumber\\
& +o(b'^2)+o({\rm tr}(B^tB)),
\end{eqnarray}
\begin{eqnarray}
\label{eq:list4}
& E[\hat{g}_{2,{\rm N}}] = p({\bf s}_{obs})h(\epsilon|{\bf s}_{obs}) +\frac{1}{2}b'^2 \mu_2(\tilde{K}) h_{\epsilon\epsilon}(\epsilon|{\bf s}_{obs}) p({\bf s}_{obs}) \nonumber \\
& + \mu_2(K)[h_{\bf s}(\epsilon|{\bf s})^t_{|{\bf s}={\bf s}_{obs}} BB^t p_{\bf s}({\bf s}_{obs})  +\frac{1}{2}h(\epsilon|{\bf s}_{obs}){\rm tr}(BB^t p_{\bf ss}({\bf s}_{obs}))  \nonumber\\
& + \frac{1}{2}p({\bf s}_{obs}){\rm tr}(BB^t h_{\bf ss}(\epsilon|{\bf s})_{|{\bf s}={\bf s}_{obs}}) +o(b'^2)+o({\rm tr}(B^tB)),
\end{eqnarray}
\begin{equation}
\label{eq:list5}
Var[\hat{g}_{\rm D}] = \frac{R(K) p({\bf s}_{obs})} {n|B|} + O(\frac{1}{n})+O(\frac{{\rm tr}(BB^t)}{n|B|}),
\end{equation}
\begin{equation}
\label{eq:list6}
Var[\hat{g}_{j,{\rm N}}] = \frac{R(K)R(\tilde{K})g(\theta|{\bf s}_{obs}) p({\bf s}_{obs})} {nb'|B|} + O(\frac{1}{n})+O(\frac{{\rm tr}(BB^t)}{nb'|B|})+O(\frac{b'}{n|B|}),\quad j=0,1,2,\\
\end{equation}
\begin{equation}
\label{eq:list7}
Cov[\hat{g}_{j,{\rm N}},\hat{g}_{\rm D}]=\frac{R(K)p({\bf s}_{obs}) g(\theta|{\bf s}_{obs})} {n|B|} +O(\frac{1}{n}), \quad j=0,1,2.
\end{equation}

\end{lemme}

\proof
See the Supplemental Material available online
\eop

Theorem \ref{th:reject_beaumont1} is a particular case of the following theorem that gives the bias and variance of the three estimators of the partial posterior distribution for a general nonsingular bandwidth matrix $B$.

\begin{theorem}
Assume that $B$ is a non-singular bandwidth matrix and assume that conditions (A1)-(A5) holds, then the bias of $\hat{g}_j$, $j=0,1,2$, is given by 

\label{th:reject_general}
\begin{equation}
\label{eqn:bias_rej_general}
E[\hat{g}_j(\theta|{\bf s}_{obs})-g(\theta|{\bf s}_{obs})]  =  D_1{b'}^2+D_{2,j}+O_P(({\rm tr}(B^tB)+b'^2)^2)+O_P(\frac{1}{n|B|}),  \; j=0,1,2,
\end{equation}

with

$$
D_1=C_1=\frac{\mu_2(\tilde{K})g_{\theta \theta}(\theta|{\bf s}_{obs})}{2},
$$
$$
D_{2,0}=\mu_2(K)\left(\frac{g_{\bf s}(\theta|{\bf s})^t_{|{\bf s}={\bf s}_{obs}} BB^t p_{\bf s}({\bf s}_{obs}) } {p({\bf s}_{obs})}+\frac{{\rm tr}(BB^t g_{\bf ss}(\theta|{\bf s})_{|{\bf s=s_{obs}}})}{2}\right), 
$$
$$
D_{2,1}=\mu_2(K) \left(
\frac{h_{\bf s}(\epsilon|{\bf s})^t_{|{\bf s}={\bf s}_{obs}} BB^t p_{\bf s}({\bf s}_{obs}) } {p({\bf s}_{obs})}+
\frac{{\rm tr}(BB^t h_{\bf ss}(\epsilon|{\bf s})_{|{\bf s}={\bf s}_{obs}})}{2}-
\frac{ h_{\epsilon}(\epsilon|{\bf s}_{obs}) {\rm tr}(BB^t m_{\bf ss})}{2}
\right),
$$
and
$$
D_{2,2}=\mu_2(K) \left(
\frac{h_{\bf s}(\epsilon|{\bf s})^t_{|{\bf s}={\bf s}_{obs}} BB^t p_{\bf s}({\bf s}_{obs}) } {p({\bf s}_{obs})}+
\frac{{\rm tr}(BB^t h_{\bf ss}(\epsilon|{\bf s})_{|{\bf s}={\bf s}_{obs}})}{2}
\right),
$$

The variance of the estimators $\hat{g}_j$, $j=0,1,2$, is given by

\begin{equation}
\label{eqn:var_general}
Var[\hat{g}_j(\theta|{\bf s}_{obs})]=\frac{R(K) R(\tilde{K}) g(\theta|{\bf s}_{obs})} {p({\bf s}_{obs})n|B|b'}(1+o_P(1)).
\end{equation}

\end{theorem}

\proof

Theorem \ref{th:reject_general} is a consequence of Lemma \ref{lem:linearize} and  \ref{lem:list}. Taking expectations on both sides of equation (\ref{eq:linearize}), we find that

\begin{equation}
\label{eq:aux1}
E[\hat{g}_j(\theta|{\bf s}_{obs}])=\frac{E[\hat{g}_{j,{\rm N}}]}{E[\hat{g}_{\rm D}]}+O_P\left[{\rm Cov}(g_{j,{\rm N}},\hat{g}_{\rm D})+{\rm Var}(\hat{g}_{\rm D})\right].
\end{equation}

Using a Taylor expansion, and the equations (\ref{eq:list1})-(\ref{eq:list4}), (\ref{eq:list5}), and (\ref{eq:list7}) given in Lemma \ref{lem:list}, we find the bias of the estimators given in equation (\ref{eqn:bias_rej_general}).

For the computation of the variance, we find from equation (\ref{eq:linearize}) and (\ref{eq:aux1}) that

\begin{equation}
\label{eq:aux2}
\hat{g}_j(\theta|{\bf s}_{obs})-E[\hat{g}_j(\theta|{\bf s}_{obs})]=\frac{\hat{g}_{j,{\rm N}}-E[\hat{g}_{j,{\rm N}}]}{E[\hat{g}_{\rm D}]}-
\frac{E[\hat{g}_{j,{\rm N}}](\hat{g}_{\rm D}-E[\hat{g}_{\rm D}])}{E[\hat{g}_{\rm D}]^2} +
O_P(\frac{1}{n|B|}).
\end{equation}

The order of the reminder follows from equations (\ref{eq:list5}) and (\ref{eq:list7}). Taking the expectation of the square of equation (\ref{eq:aux2}), we now find

\begin{equation}
\label{eq:aux3}
{\rm Var}[\hat{g}_j(\theta|{\bf s}_{obs}])=\frac{{\rm Var}[\hat{g}_{j,{\rm N}}]}{E[\hat{g}_{\rm D}]^2} + \frac{E[\hat{g}_{j,{\rm N}}]^2 {\rm Var}[\hat{g}_{\rm D}]} {E[\hat{g}_{\rm D}]^4} - 2 {\rm Cov}(\hat{g}_{\rm D},\hat{g}_{j,{\rm N}})\frac{E[\hat{g}_{j,{\rm N}}]}{E[\hat{g}_{\rm D}]^3}+ o_{P}(\frac{1}{n|B|b'}).
\end{equation}
The variance of the estimators given in equation (\ref{eqn:var_general}) follows from a Taylor expansion that makes use of equations (\ref{eq:list1})-(\ref{eq:list7}) given in Lemma \ref{lem:list}.

%
%
%
\eop

\section{Supplemental Materials}
\begin{description}
\item Proof of Lemma \ref{lem:list}
\end{description}

\bibliographystyle{ECA_jasa}
\bibliography{bibjasa}

\end{document}